\documentclass[12pt]{article}

\makeatletter
 
\@addtoreset{equation}{section}
\makeatother

\newcommand{\hs}{\hspace}
\newcommand{\ts}{\hspace*}
\newcommand{\vs}{\vspace}
\newcommand{\e}{\enskip}
\newcommand{\q}{\quad}

\newcommand{\dps}{\displaystyle}
\newcommand{\f}{\frac}

\newcommand{\Phat}{\hat{{\cal P}}}
\newcommand{\Qhat}{\hat{\Omega}}

\newcommand{\xipp}[1]{\hat{\xi}^{\mbtn{#1}(+)}}
\newcommand{\ximm}[1]{\hat{\xi}^{\mbtn{#1}(-)}}

\newcommand{\pimm}[1]{\hat{\pi}^{\mbtn{#1}(-)}}

\newcommand{\ximms}[2]{\hat{\xi}^{\mbtn{#1}(-)}_{#2}}

\newcommand{\pimms}[2]{\hat{\pi}^{\mbtn{#1}(-)}_{#2}}

\newcommand{\Zmms}[2]{\hat{Z}^{\mbtn{#1}(-)}_{#2}}

\newcommand{\mcK}{\mathcal{K}}
\newcommand{\mcC}{\mathcal{C}}
\newcommand{\mcA}{\mathcal{A}}

\newcommand{\mcS}{\mathcal{S}}

\newcommand{\mcG}{\mathcal{G}}

\newcommand{\commu}[2]{[#1,#2]}
\newcommand{\commut}[2]{[#1,\e #2]}
\newcommand{\symmp}[2]{\{#1,#2\}}

\newcommand{\mbtn}[1]{\mbox{{\tiny #1}}}

\begin{document}

\begin{center}
{\bf \Large Unified Description of Quantum Mechanics on a Curved Space
}\vs{12pt}\\
M. Nakamura\footnote{\label{*}E-mail:mnakamur@hm.tokoha-u.ac.jp}
\vs{12pt}\\
{\it Research Institute, Hamamatsu Campus, Tokoha University, Miyakoda-cho 1230, 
Kita-ku, Hamamastu-shi, Shizuoka 431-2102, Japan}
\end{center}

\begin{abstract}
Starting with the first-order singular Lagrangian, the problem of the quantization of a dynamical system constrained to a submanifold embedded in the higher-dimensional Euclidean space is investigated
within the framework of operatorial quantization formalism. Through the projection operator method (POM) with the constraint star-products, it is shown that both of the constraint quantum  system with the usual constraint and that with the derivative-type constraint are naturally constructed from one Lagarangian. It is  proved that the system with the usual constraint is the sub-system of that with the derivative-type one. Furthermore, the quantization of the dynamical system subject to both of the usual constraint and the derivative-type one is investigated by the POM, and the quantum corrections in the resultant Hamiltonians are discussed.
\end{abstract}


\section{Introduction}	

\ts{12pt}The problem of the quantization of a dynamical system constrained to a submanifold   embedded in the higher-dimensional Euclidean space  has been extensively investigated as one of the quantum theories on a curved space until now\cite{A1,A2,A3,A4}. In order to avoid the unnecessary troublesomeness, the submanifold $M^{N-1}$ specified by $G(x)=0$ ($G(x)\in {\it C}^{\infty}$) in an $N$-dimensional Euclidean space $R^N$ has been considered in many studies, where $x=(x^1,\cdots,x^i,\cdots,x^N)\in R^N$. As the dynamical system constrained to $M^{N-1}$, then, the dynamical system subject to the constraint $G(x)=0$ and that subject to $\dot{G}(x)=0$\footnote{We shall call the constraint $G(x)=0$ the {\it static} constraint, and $\dot{G}(x)=0$, the {\it dynamical} one.} have been presented with the Lagrangians $L=L_0+\mu G(x)$, =$L_0+\mu \dot{G}(x)$, where $L_0$ is the usual  second-order nonsingular Lagrangian and $\mu$, an additional dynamical variable\cite{A3,A4}. In this paper, we shall call the system subject to $G(x)=0$ the {\it static} constraint system, and the system subject to $\dot{G}(x)=0$, the {\it dynamical} constraint system.\\
\ts{12pt}In the quantization of constraint systems, alternative approach of the first-order singular Lagrangian formalism has been proposed by Faddeev and Jackiw\cite{A5}. Starting with the first-order singular Lagrangian containing the term associated to the {\it dynamical} constraint, in this paper, we shall accomplish the quantization of the system constrained to $M^{N-1}$ within the framework of operatorial quantization formalism\cite{A6}. Through the projection operator method(POM) with the constraint star-products\cite{A7}, then, it is shown that the resultant constraint quantum systems corresponding to these two constraints, the {\it static} constraint quantum system and the {\it dynamical} constraint quantum system, are naturally constructed with one Lagrangian, and it is proved, with the alternative approach to Ref.\cite{A3}, that the {\it static} constraint quantum system is the sub-system of the {\it dynamical} constraint quantum system . Furthermore, we shall discuss the quantum correction terms due to the projections of operators and the re-ordering of operators in the resultant Hamiltonian.\\
\ts{12pt}The quantization of the dynamical sytems subject to both of the {\it static} constraint and the {\it dynamical} one has been investigated through, for example, the St\"{u}ckelberg formalism\cite{A10}. Starting with the first-order singular Lagrangian containing these two consraint terms, we shall 
quantize the dynamical system with these two kinds of contraints by the POM, and it is shown that the resultant constraint quantum system is included in the constraint quantum system with the {\it dynamical} constraint. \\
\ts{12pt}The program of this paper is the following. In Sect.2, we propose the Lagrangian with the {\it dynamical} constraint and construct the initial unconstraint quantum system, which we denote $\mcS$. In Sect.3, we construct the the {\it static} constraint quantum system, which we denote $\mcS^*_{\mbtn{I}}$,  the {\it dynamical} constraint quantum system, which is denoted by $\mcS^*_{\mbtn{II}}$, and the constraint quantum system with the {\it static} and {\it dynamical} constraints, which is denoted by $\mcS^*_{\mbtn{III}}$. Then, it is proved that $\mcS^*_{\mbtn{I}}$ is the sub-system of $\mcS^*_{\mbtn{II}}$, and $\mcS^*_{\mbtn{III}}$ is included in $\mcS^*_{\mbtn{II}}$. Furthermore, the quantum correction terms in the final Hamiltonians are discussed. In Sect.4, the discussion and the some concluding remarks are given.

\section{Initial Hamiltonian System}

\ts{12pt}Consider the dynamical system described by the first-order singular Lagrangian $L$
$$
L=L(x,\dot{x},v,\dot{v},\lambda,\dot{\lambda})
=\dot{x}^iv_i
-\lambda\dot{G}(x)-\f12v_iv_i-V(x),
\eqno{(2.1)}
$$
where $x=(x^1,\cdots,x^i\cdots,x^N),v=(v_1,\cdots,v_i,\cdots,v_N)$ and $\dot{G}(x)=\dot{x}^iG_i(x)$\footnote{Here, $G_{i\cdots j}(x)$ stands for $\partial^x_i\cdots\partial^x_jG(x)$ with $\partial^x_i=\partial/\partial x^i.$}.\\
\ts{12pt}Following the canonical quantization formulation for constraint systems\cite{A8,A9}, then, the initial unconstraint quantum system $\mcS=(\mcC,\mcA(\mcC),H(\mcC),\mcK)$ is obtained as follows: \vs{12pt}\\
{\bf \boldmath i) Initial canonically conjugate set $\mcC$}
$$ 
\ts{-72pt}\mcC=\{(x^i,p^x_i),(v_i,p_v^i),(\lambda,p_{\lambda})|i=1,\cdots,N\},
\eqno{(2.2)}
$$
which obeys the commutator algebra $\mcA(\mcC)$:
$$
\commut{x^i}{p^x_j}=i\hbar\delta^i_j,\hs{12pt}\commut{v_i}{p_v^j}=i\hbar\delta_i^j,\hs{12pt}
\commut{\lambda}{\pi_{\lambda}}=i\hbar,\hs{12pt}
\mbox{(the others)}=0,
\eqno{(2.3)}
$$
{\bf \boldmath ii) Initial Hamiltonian $H$}
$$
H=\{\mu^i_{(1)},\phi^{(1)}_i\}+\{\mu^i_{(2)},\phi^{(2)}_i\}+\{\mu_{(3)},\phi^{(3)}\}+\f12v_iv_i+V(x),
\eqno{(2.4)}
$$
where $\phi^{(n)}$, $(n=1,\cdots, 3)$ are the constraint operators corresponding to the primary constraints $\phi^{(n)}\approx 0$ due to the singularity of the Lagrangian $L$, which are given by
$$
\begin{array}{lcl}
\phi^{\mbtn{(1)}}_i&=&p^x_i-v_i+\lambda G_i(x),\vs{12pt}\\
\phi^{\mbtn{(2)}}_i&=&p_v^i,\vs{12pt}\\
\phi^{\mbtn{(3)}}&=&p_{\lambda},
\end{array}
\eqno{(2.5)}
$$
and $\mu^i_{(1)}$, $\mu^i_{(2)}$ and $\mu_{(3)}$ are the Lagrange multiplier operators.
\vs{12pt}\\
{\bf \boldmath iii) Consistent set of constraints}
\vs{6pt}\\
\ts{12pt}From the consistency conditions for the time evolusions of constraint operators, the consistent set of constraints, $\mcK$, is given by
$$
\mcK=\{\phi^{\mbtn{(1)}}_i,\phi^{\mbtn{(2)}}_i,\phi^{\mbtn{(3)}},\psi^{\mbtn{(1)}}\},
\eqno{(2.6)}
$$
where  
$$
\psi^{\mbtn{(1)}}=G_i(x)v_i,
\eqno{(2.7)}
$$
which is the constraint operator corresponding to the secondary constraint. \\
\ts{12pt}Then, the Lagrange multiplier operators are determined as
$$
\begin{array}{lcl}
\mu_{\mbtn{(1)}}^i&=&v_i,\vs{12pt}\\
\mu_{\mbtn{(2)}}^i&=&-\mcG^{-1}(x)G_i(x)G_{kl}(x)v_kv_l-P_{ij}V_j(x),\vs{12pt}\\
\mu_{\mbtn{(3)}}&=&\mcG^{-1}(x)(V_i(x)G_i(x)-G_{ij}(x)v_iv_j),
\end{array}
\eqno{(2.8)}
$$
where
$$
\begin{array}{l}
\mcG(x)=G_i(x)G_i(x),\vs{12pt}\\
P_{ij}=\delta_{ij}-\mcG^{-1}(x)G_i(x)G_j(x),
\end{array}
\eqno{(2.9)}
$$
which satisfies
$$
P_{ij}P_{jk}=P_{ik},\hs{36pt}P_{ij}G_j(x)=G_i(x)P_{ij}=0.
\eqno{(2.10)}
$$
{\bf \boldmath iv) Commutator algebra of $\mcK$}
\vs{6pt}\\
\ts{12pt}The constraint set $\mcK$ obeys the commutator algebra $\mcA(\mcK)$:
$$
\begin{array}{lcll}
\mcA(\mcK)&:&\commu{\phi^{\mbtn{(1)}}_i}{\phi^{\mbtn{(2)}}_j}=-i\hbar\delta_{ij},&\vs{12pt}\\

& &\commu{\phi^{\mbtn{(1)}}_i}{\phi^{\mbtn{(3)}}}=i\hbar G_i(x),&\vs{12pt}\\

& &\commut{\psi^{\mbtn{(1)}}}{\phi^{\mbtn{(1)}}_i}=i\hbar G_{ij}(x)v_j,&\vs{12pt}\\

& &\commut{\psi^{\mbtn{(1)}}}{\phi^{\mbtn{(2)}}_i}=i\hbar G_i(x),&\mbox{(the others)}=0.\vs{12pt}\\
\end{array}
\eqno{(2.11)}
$$

\section{The Constraint Quantum System $\mcS^*$}

\ts{12pt}Starting with the initial system $\mcS$, we shall construct the constraint quantum system $\mcS^*$, which satisfies $\mcK=0$, through the star-product quantization formalism of POM\cite{A7}.\\ 
\ts{12pt}For this purpose, we first classify $\mcK$ into the following two subsets  :
$$
\mcK=\mcK^{(\mbtn{A})}\oplus\mcK^{\mbtn{(B)}}\q \mbox{with}\q \mcK^{(\mbtn{A})}=\{\phi^{\mbtn{(1)}},\phi^{\mbtn{(2)}}\},\hs{12pt}\mcK^{\mbtn{(B)}}=\{\phi^{\mbtn{(3)}},\psi^{\mbtn{(1)}}\}.
\eqno{(3.1)}
$$
As well as the Dirac bracket formalism, the POM satisfies the {\it iterative} property\cite{A9}. Terefore, the constraint quantum system $\mcS^*$ can be constructed through the successive projections of $\mcS$\cite{A7}.

\subsection{Successive Projections of $\mcS$}
\ts{12pt}From the structure of the commutator algebra (2.10), the successive projections of $\mcS$ can be uniquely carried out through the following diagram: 
$$
\Phat^{\mbtn{(1)}}\mcK^{\mbtn{(1)}}=0\rightarrow\Phat^{\mbtn{(2)}}\mcK^{\mbtn{(2)}}=0,
\eqno{(3.2)}
$$
where 
$$
\mcK^{\mbtn{(1)}}=\mcK^{\mbtn{(A)}},\hs{12pt}\mcK^{\mbtn{(2)}}=\mcK^{\mbtn{(B)}},
\eqno{(3.3)}
$$ 
and $\Phat^{\mbtn{(n)}}$ is the projection operator associated to the subset $\mcK^{\mbtn{(n)}}$, that is, $\Phat^{\mbtn{(n)}}\mcK^{\mbtn{(n)}}=0$ $(n=1,2)$.  Then, the successive projections of the operators of the system by $\Phat^{(n)}$ $(n=1,2)$ are carried out through the program designated by the following diagram :
$$
\mcC\stackrel{\Phat^{(1)}}{\longrightarrow}\mcC^{(1)}\stackrel{\Phat^{(2)}}{\longrightarrow}\mcC^{(2)},
\eqno{(3.4)}
$$
where 
$$
\mcC^{(n)}=\Phat^{(n)}\mcC^{(n-1)}\hs{60pt}(n=1,2)
\eqno{(3.5)}
$$
with $\mcC^{(0)}=\mcC$, which satisfy the projection conditions
$$
\mcK^{(n)}(\mcC^{(n)})=0\hs{60pt}(n=1,2).
\eqno{(3.6)}
$$
Then, the ACCS (associated canonically conjugate set) $Z^{(n)}$\cite{A7} for the subsets $\mcK^{(n)}$ $(n=1,2)$ consist of the operators in $\mcC^{(n-1)}$,
$$
Z^{(n)}=Z^{(n)}(\mcC^{(n-1)}).
\eqno{(3.7)}
$$
From (3.7), therefore, the projection operators $\Phat^{(n)}$ are also represented as 
$$
\Phat^{(n)}=\Phat^{(n)}(\mcC^{(n-1)})\hs{60pt}(n=1,2).
\eqno{(3.8)}
$$ 
\vs{12pt}\\
\ts{12pt}Now, The ACCS $Z^{\mbtn{(n)}}_{\alpha}$ of the projection operators $\Phat^{(n)}$ ($n=1,2$) are given, respectively, as follows: 
$$
\begin{array}{clcl}

(1)&Z^{\mbtn{(1)}}_{\alpha}=Z^{\mbtn{(1)}}_{\alpha}(\mcC)&=&\left\{\begin{array}{rcll}\xi^{\mbtn{(1)}}_i&=&-\phi^{\mbtn{(1)}}_i=v_i-p^x_i-\lambda G_i(x)&(\alpha=i),\vs{12pt}\\
\pi^{\mbtn{(1)}}_i&=&\phi^{\mbtn{(2)}}_i=p_v^i&(\alpha=i+N),
\end{array}\right.
\vs{12pt}\\
& & &\hs{24pt}(\alpha=1,\cdots,2N\q;\q i=1,\cdots,N),
\vs{24pt}\\

(2)&Z^{\mbtn{(2)}}=Z^{\mbtn{(2)}}(\mcC^{(1)})&=&\left\{\begin{array}{rcll}\xi^{\mbtn{(2)}}&=&\symmp{\mcG^{-1}(x)}{\psi^{\mbtn{(1)}}}& \vs{6pt}\\
&=&\symmp{\mcG^{-1}(x)G_i(x)}{v_i}&\hs{24pt}(\alpha=1),\vs{12pt}\\
\pi^{\mbtn{(2)}}&=&\phi^{\mbtn{(3)}}=p_{\lambda}&\hs{24pt}(\alpha=2).
\end{array}\right.

\end{array}
\eqno{(3.9)}
$$
\ts{12pt}The {\it hyper}-operators $\Qhat^{\mbtn{(n)}}_{\eta\zeta}$ for $\mcK^{\mbtn{(n)}}$ $(n=1,2)$ in the star-product formulation are given as follows:\footnote{For any operator $O$, $\hat{O}^{(-)}\bullet=\f1{i\hbar}\commut{O}{\bullet}$ and $\hat{O}^{+}\bullet=\symmp{O}{\bullet}=\f12(O\bullet+\bullet O).$}
$$
\begin{array}{l}
\Qhat^{\mbtn{(1)}}_{\eta\zeta}=J^{\alpha\beta}\Zmms{(1)}{\alpha}(\eta)\Zmms{(1)}{\beta}(\zeta)=\ximms{(1)}{i}(\eta)\pimms{(1)}{i}(\zeta)-\pimms{(1)}{i}(\eta)\ximms{(1)}{i}(\zeta),\vs{12pt}\\
\Qhat^{\mbtn{(2)}}_{\eta\zeta}=J^{\alpha\beta}\Zmms{(2)}{\alpha}(\eta)\Zmms{(2)}{\beta}(\zeta)=\ximms{(2)}{}(\eta)\pimms{(2)}{}(\zeta)-\pimms{(2)}{}(\eta)\ximms{(2)}{}(\zeta).
\end{array}
\eqno{(3.10)}
$$
The operations of $\ximm{(n)}$ and $\pimm{(n)}$ on $\mcC^{\mbtn{(n-1)}}$ are presented in Appendix A.

\subsubsection{Projected CCS $\mcC^{\mbtn{(2)}}$ and commutator algebra of $\mcC^{\mbtn{(2)}}$}

\ts{12pt}Let $\Phat$ be $\Phat=\Phat^{\mbtn{(2)}}\Phat^{\mbtn{(1)}}$. From the projection conditions (3.6), then, $\mcC^{\mbtn{(2)}}$ becomes as follows:
$$
\begin{array}{rcl}
\mcC^{\mbtn{(2)}}&=&\Phat\mcC=\mcC\{(\Phat x,\Phat p^x),(\Phat v,\Phat p_v),(\Phat\lambda,\Phat\pi_{\lambda})\}\vs{6pt}\\
&=&\{x^i,p^x_i,v_i,\lambda|i=1,\cdots,N\},
\end{array}
\eqno{(3.11)}
$$
where the projection conditions are represented as
$$
\begin{array}{l}
\Phat\phi^{\mbtn{(1)}}_i=p^x_i-v_i+\symmp{\lambda}{G_i(x)}=0,\vs{12pt}\\
\Phat\phi^{\mbtn{(2)}}_i=p_v^i=0,\hs{24pt}\Phat\phi^{\mbtn{(3)}}=p_{\lambda}=0,\vs{12pt}\\
\Phat\psi^{\mbtn{(1)}}=\symmp{G_i(x)}{v_i}=0.
\end{array}
\eqno{(3.12)}
$$
Then, the commutator algebra $ \mcA(\mcC^{\mbtn{(2)}})$ is represented as 
$$
\begin{array}{l}
\commut{x^i}{p^x_j}=i\hbar\delta^i_j,\hs{24pt}\commut{x^i}{v_j}=i\hbar P_{ij}(x),\hs{24pt}\commut{\lambda}{x^i}=i\hbar\mcG^{-1}(x)G_i(x),\vs{12pt}\\

\commut{v_i}{v_j}=i\hbar\symmp{\mcG^{-1}(G_{ik}G_j(x)-G_i(x)G_{jk}(x))}{v_k},\vs{12pt}\\

\commut{v_i}{p^x_j}=i\hbar\symmp{\lambda}{P_{ik}G_{kj(x)}}-i\hbar\symmp{\mcG^{-1}(x)G_i(x)G_{jk}(x)}{v_k},\vs{12pt}\\

\commut{p^x_i}{\lambda}=i\hbar\symmp{\mcG^{-1}(x)G_{ik}(x)}{v_k}+i\hbar\symmp{\lambda}{\mcG^{-1}(x)G_{ik}(x)G_k(x)},\vs{12pt}\\

\commut{v_i}{\lambda}=i\hbar \symmp{\mcG^{-1}(x)G_{ik}}{v_k},\hs{48pt}\mbox{(the others)}=0,

\end{array}
\eqno{(3.13)}
$$

\subsubsection{Projected Hamiltonian $H^{\mbtn{(2)}}$}

\ts{12pt}According to the projection formulas of symmetrized product\cite{A7}, the projected Hamiltonian $H^{\mbtn{(2)}}$ is obtained in the following way:
$$
H^{\mbtn{(2)}}=\Phat H=\Phat^{\mbtn{(2)}}\Phat^{\mbtn{(1)}}H=\dps{\f12}\symmp{v_i}{v_i}+V(x)+V_{\mbtn{C}}^{\mbtn{(1)}}(x)+V_{\mbtn{C}}^{\mbtn{(2)}}(x),
\eqno{(3.14)}
$$
where $V_{\mbtn{C}}^{\mbtn{(1)}}(x)$ is the quantum correction term due to the projection of $\f12\symmp{v_i}{v_i}$,
$$
V_{\mbtn{C}}^{\mbtn{(1)}}(x)=\f{\hbar^2}8\mcG^{-2}(x)G_{ik}(x)G_k(x)G_{il}(x)G_l(x)-\dps{\f{\hbar^2}4}\mcG^{-1}(x)G_{ik}(x)P_{kl}G_{li},
\eqno{(3.15)}
$$
and  
$V_{\mbtn{C}}^{\mbtn{(2)}}(x)$, that due to the projection of $\symmp{\mu_{\mbtn{(3)}}}{\phi^{\mbtn{(3)}}}$,
$$
V_{\mbtn{C}}^{\mbtn{(2)}}(x)=\f{\hbar^2}4\mcG^{-1}(x)G_{ij}(x)G_{ij}(x).
\eqno{(3.16)}
$$
Substituting $P_{ij}=\delta_{ij}-\mcG^{-1}(x)G_i(x)G_j(x)$ into (3.15), then, one obtains
$$
H^{\mbtn{(2)}}=\dps{\f12}\symmp{v_i}{v_i}+V(x)+\dps{\f{3\hbar^2}8}\mcG^{-2}(x)G_{ik}(x)G_k(x)G_{il}(x)G_l(x).
\eqno{(3.17)}
$$
\\
\ts{12pt}Thus, we have obtained the final projected quantum system due to the successive projection $\Phat=\Phat^{\mbtn{(2)}}\Phat^{\mbtn{(1)}}$,
$$
\mcS^{\mbtn{(2)}}=(\mcC^{\mbtn{(2)}},\mcA(\mcC^{\mbtn{(2)}}),H^{\mbtn{(2)}}).
\eqno{(3.18)}
$$

\subsection{Constraint Quantum System $\mcS^*$}

\ts{12pt}From the  structure of the commutator algebra $\mcA(\mcC^{\mbtn{(2)}})$, we can consider two  constraint quantum systems, one of which is represented in terms of $\{x, v\}$, and the other, in terms of $\{x,p^x\}$. 

\subsubsection{Constraint Quantum System $\mcS^*_{\mbtn{I}}$}

\ts{12pt}Let $\{x,v\}$ be the CCS $\mcC^*_{\mbtn{I}}$ in $\mcS^*_{\mbtn{I}}$,
$$
\mcC^*_{\mbtn{I}}=\{x^i,v_i|i=1,\cdots,N\}.
\eqno{(3.19)}
$$
From the projection conditions (3.12), $\mcC^*_{\mbtn{I}}$ satisfies  
$$
\begin{array}{l}
\symmp{P_{ij}}{p^x_j}=v_i,\vs{6pt}\\
\lambda=\symmp{\mcG^{-1}(x)G_i(x)}{p_i},\vs{6pt}\\
p_v^i=0,\hs{36pt}p_{\lambda}=0.
\end{array}
\eqno{(3.20)}
$$
\ts{12pt}Let $n_i(x)$ ,$n_{i;k}(x)$ be $n_i(x)=\mcG^{-1/2}G_i(x)$, $n_{i;k}(x)=\partial^x_kn_i(x)$, respectively. Then, the commutator algebra $\mcA(\mcC^*_{\mbtn{I}})$ is defined as follows:

$$
\begin{array}{lcl}
\mcA(\mcC^*_{\mbtn{I}})&:&\commut{x^i}{x^j}=0,\vs{12pt}\\
& &\commut{x^i}{v_j}=i\hbar P_{ij}=i\hbar(\delta_{ij}-n_i(x)n_j(x)),\vs{12pt}\\
& &\commut{v_i}{v_j}=i\hbar\symmp{n_{i;k}(x)n_j(x)-n_i(x)n_{j;k}(x)}{v_k},
\end{array}
\eqno{(3.21)}
$$
which is just equivalent to the commutator algebra of the {\it static} constraint quantum system\cite{A1,A3,A4}. From the projection conditions (3.20), however, $\mcA(\mcC^*_{\mbtn{I}})$ does not reproduce the commutator algebra with respect to $p^x_i$ and $\lambda$.\\
\ts{12pt}The Hamiltonian $H^*_{\mbtn{I}}$ in $\mcS^*_{\mbtn{I}}$ is given by the projected Hamiltonian $H^{\mbtn{(2)}}$, which is also expressed in the following form: 
$$
H^*_{\mbtn{I}}=H^{\mbtn{(2)}}=\dps{\f12}\symmp{v_i}{v_i}+V(x)+\dps{\f{3}{32}}\hbar^2\mcG^{-2}(x)\mcG_{;i}(x)\mcG_{;i}(x),
\eqno{(3.22)}
$$
where
$$
\mcG_{;i}(x)=\partial^x_i\mcG(x).
\eqno{(3.23)}
$$
\ts{12pt}Thus, the quantum system $\mcS^*_{\mbtn{I}}$ is defined with
$$
\mcS^*_{\mbtn{I}}=(\mcC^*_{\mbtn{I}},\mcA(\mcC^*_{\mbtn{I}}),H^*_{\mbtn{I}}).
\eqno{(3.24)}
$$ 
Following $\mcA(\mcC^*_{\mbtn{I}})$, then, it is obvious that $\dot{G}(x)=0$, that is,
$$
\dot{G}(x)=\f1{i\hbar}\commut{G(x)}{H^*_{\mbtn{I}}}=\symmp{P_{ij}G_j(x)}{v_i}=0.
\eqno{(3.25)}
$$
In the quantum system $\mcS^*_{\mbtn{I}}$, thus, $G(x)$ is the constant of motion and $G(x)=0$ is conserved through the time evolusion of system under the operatorial formalisn. Because of the projection conditions (3.20), on the other hand, $\mcA(\mcC^*_{\mbtn{I}})$ is not {\it complete} with respect to $p^x_i,\lambda$. The commutator algerbra of these operators is defined by $\mcA(\mcC^{\mbtn{(2)}})$, which shows that $\mcC^*_{\mbtn{I}}$ is the subset of $\mcC^{\mbtn{(2)}}$,
$$
\mcC^*_{\mbtn{I}}\subset\mcC^{\mbtn{(2)}},
\eqno{(3.26)}
$$
and, therefore, $\mcS^*_{\mbtn{I}}$ is the sub-system of $\mcS^{\mbtn{(2)}}$,
$$
\mcS^*_{\mbtn{I}}\subset \mcS^{\mbtn{(2)}}.
\eqno{(3.27)}
$$
\subsubsection{Constraint Quantum System $\mcS^*_{\mbtn{II}}$}

\ts{12pt}The alternative final CCS is given by $\{x,p^x\}$:
$$
\mcC^*_{\mbtn{II}}=\{x^i,p^x_i|i=1,\cdots,N\}.
\eqno{(3.28)}
$$
From $\mcA(\mcC^{\mbtn{(2)}})$, the commutator algebra $\mcA(\mcC^*_{\mbtn{II}})$ is defined by the canonically conjugate commutation relations
$$
\mcA(\mcC^*_{\mbtn{II}}):\commut{x^i}{x^j}=0,\hs{24pt}\commut{x^i}{p^x_j}=i\hbar\delta^i_j,\hs{24pt}\commut{p^x_i}{p^x_j}=0,
\eqno{(3.29)}
$$
which is just identical to the commutator algebara in the {\it dynamical} constraint quantum system\cite{A3,A4}. 
According to the projection conditions (3.12), then, $v_i$, $\lambda$ are represented in terms of $x^i,p^x_i$ as follows:
$$
v_i=\symmp{P_{ij}}{p^x_i},\hs{24pt}\lambda=\symmp{\mcG^{-1}(x)G_i(x)}{p^x_i},
\eqno{(3.30)}
$$
and the commutator algebra with respect to $v_i$ and $\lambda$ is completely reproduced by $\mcA(\mcC^*_{\mbtn{II}})$. 
Therefore, $\mcC^{\mbtn{(2)}}$ is represented as
$$
\mcC^{\mbtn{(2)}}=\{x,p^x,v,\lambda\}=\{x,p^x,v(x,p^x),\lambda(x,p^x)\},
\eqno{(3.31)}
$$
which shows that $\mcC^*_{\mbtn{II}}$ is just equivalent to $\mcC^{\mbtn{(2)}}$,
$$
\mcC^{\mbtn{(2)}}=\mcC^*_{\mbtn{II}}.
\eqno{(3.32)}
$$
\ts{12pt}From (3.22) and (3.30), the Hamiltonian $H^*_{\mbtn{II}}$ is represented as
$$
H^*_{\mbtn{II}}=\f12\symmp{\symmp{P_{ik}}{p^x_k}}{\symmp{P_{il}}{p^x_l}}+V(x)+\dps{\f{3}{32}}\hbar^2\mcG^{-2}(x)\mcG_{;i}(x)\mcG_{;i}(x),
\eqno{(3.33)}
$$
which contains the products of operators noncommutatable with each other.  So, the re-ordering of operators also yields the quantum correction terms in $H^*_{\mbtn{II}}$.  \\
\ts{12pt}We shall rewrite $\f12\symmp{\symmp{P_{ik}}{p^x_k}}{\symmp{P_{il}}{p^x_l}}$ as follows:
$$
\f12\symmp{\symmp{P_{ik}}{p^x_k}}{\symmp{P_{il}}{p^x_l}}=\f12p^x_iP_{ij}p^x_j+U_c(x),
\eqno{(3.34)}
$$
where $U_c(x)$ is the quantum correction term associated to the re-ordering of operators, 
$$
U_c(x)=\f{\hbar^2}4(n_{k;kl}(x)n_l(x)+n_{k;l}(x)n_{l;k}(x))+\f{\hbar^2}8(n_{k;k}(x)n_{l;l}(x)+n_k(x)n_{i;k}(x)n_{i;l}(x)n_l(x)).
\eqno{(3.35)}
$$
Then, $H^*_{\mbtn{II}}$ is rewritten in the following form: 
$$
H^*_{\mbtn{II}}=\f12p^x_iP_{ij}p^x_j+V(x)+\f3{32}\hbar^2\mcG^{-2}(x)\mcG_{;i}(x)\mcG_{;i}(x)+U_c(x).
\eqno{(3.36)}
$$
Thus, the quantum system $\mcS^*_{\mbtn{II}}$ is given by
$$
\mcS^*_{\mbtn{II}}=(\mcC^*_{\mbtn{II}},\mcA(\mcS^*_{\mbtn{II}}),H^*_{\mbtn{II}}),
\eqno{(3.37)}
$$
and it is easily shown that $\dot{G}(x)=0$  holds in $\mcS^*_{\mbtn{II}}$ also, that is,
$$
\dot{G}(x)=\f1{i\hbar}\commut{G(x)}{H^*_{\mbtn{II}}}=\symmp{p^x_i}{P_{ij}G_j(x)}=0.
\eqno{(3.38)}
$$

\subsection{Constraint Quantum System $\mcS^*_{\mbtn{III}}$}

\ts{12pt}Consider the dynamical system described by the first-order singular Lagrangian containing terms corresponding to the {\it static} and {\it dynamical} constraints, $L_{sd}$, 
$$L_{sd}=L(x,\dot{x},v,\dot{v},\nu,\dot{\nu},\lambda,\dot{\lambda})
=\dot{x}^iv_i-\nu G(x)-\lambda\dot{G}(x)-\f12v_iv_i-V(x).
\eqno{(3.39)}
$$ 
\subsubsection{Initial Unconstraint Quantum System}

\ts{12pt}As well as in the case of the Lagrangian (2.1), the initial unconstraint quantum system $\mcS_{sd}=(\mcC, \mcA(\mcC),H(\mcC),\mcK)$ is given as follows:\vs{6pt}\\
{\bf \boldmath i) Initial canonically conjugate set $\mcC$} 
$$
\begin{array}{l}
\mcC=\{(x^i,p^x_i),(v_i,p_v^i),(\nu,p_{\nu}),(\lambda,p_{\lambda})|i=1,\cdots,N\},\vs{6pt}\\
\mcA(\mcC):\vs{6pt}\\
\commut{x^i}{p^x_j}=i\hbar\delta^i_j,\e\commut{v_i}{p_v^j}=i\hbar\delta_i^j,\e
\commut{\nu}{p_{\nu}}=i\hbar,\e \commut{\lambda}{p_{\lambda}}=i\hbar,\e
\mbox{(the others)}=0.
\end{array}
\eqno{(3.40)}
$$
{\bf \boldmath ii) Consistent set of constraints $\mcK$}
$$
\mcK=\{\phi^{\mbtn{(1)}}_i,\phi^{\mbtn{(2)}}_i,\phi^{\mbtn{(3)}},\phi^{\mbtn{(4)}},\psi^{\mbtn{(1)}}, \psi^{\mbtn{(2)}}|i=1,\cdots,N\},
\eqno{(3.41a)}
$$
where
$$
\begin{array}{l}
\phi^{\mbtn{(1)}}_i=p^x_i-v_i+\lambda G_i(x),\e \phi^{\mbtn{(2)}}_i=p_v^i,\e \phi^{\mbtn{(3)}}=p_{\nu},
\e \phi^{\mbtn{(4)}}=p_{\lambda},\vs{6pt}\\
\psi^{\mbtn{(1)}}=G(x),\e \psi^{\mbtn{(2)}}=G_i(x)v_i,
\end{array}
\eqno{(3.41b)}
$$
which obeys $\mcA(\mcK)$:
$$
\begin{array}{ll}
\commut{\phi^{\mbtn{(1)}}_i}{\phi^{\mbtn{(2)}}_j}=-i\hbar\delta_{ij},&\commut{\phi^{\mbtn{(1)}}_i}{\phi^{\mbtn{(4)}}}=i\hbar G_i(x),\vs{12pt}\\
\commut{\psi^{\mbtn{(1)}}}{\phi^{\mbtn{(1)}}_i}=i\hbar G_i(x)&\commut{\psi^{\mbtn{(2)}}}{\phi^{\mbtn{(1)}}_i}=i\hbar G_{ij}(x)v_j,\vs{12pt}\\
\commut{\psi^{\mbtn{(2)}}}{\phi^{\mbtn{(2)}}_i}=i\hbar G_i(x),&\mbox{(the others)}=0.
\end{array}
\eqno{(3.42)}
$$
{\bf \boldmath iii) Initial Hamiltonian $H$}
$$
H=\{\mu^i_{(1)},\phi^{(1)}_i\}+\{\mu^i_{(2)},\phi^{(2)}_i\}+\{\mu_{(3)},\phi^{(3)}\}+\{\mu_{(4)},\phi^{(4)}\}+\f12v_iv_i+V(x)+\nu G(x),
\eqno{(3.43)}
$$
where the Lagrange multiplier operators associated with $\dot{x}^i$, $\dot{v}_i$, $\dot{\nu}$ and $\dot{\lambda}$ are determined as 
$$
\begin{array}{lcl}
\mu_{\mbtn{(1)}}^i&=&v_i,\vs{12pt}\\
\mu_{\mbtn{(2)}}^i&=&-\mcG^{-1}(x)G_i(x)G_{kl}(x)v_kv_l-P_{ij}V_j(x),\vs{12pt}\\
\mu_{\mbtn{(4)}}&=&\nu+\mcG^{-1}(x)(V_i(x)G_i(x)-G_{ij}(x)v_iv_j),
\end{array}
\eqno{(3.44)}
$$
and $\mu_{\mbtn{(3)}}$ is indefinite.

\subsubsection{Successive Projections of $\mcS_{sd}$ and Constraint Quantum System $\mcS^*_{\mbtn{III}}$}

\ts{12pt}From the structure of $\mcA(\mcK)$, (3.42), the successive projections of $\mcS_{sd}$ can be carried out through two ways of the projection process designated by the following diagrams, respectively:  
$$
\{\phi^{\mbtn{(1)}},\phi^{\mbtn{(2)}}\}=0\rightarrow\{\psi^{\mbtn{(1)}},\psi^{\mbtn{(2)}}\}=0\rightarrow\{\phi^{\mbtn{(3)}},\phi^{\mbtn{(4)}}\}=0,
\eqno{\mbox{(P1)}}
$$
$$
\{\phi^{\mbtn{(1)}},\phi^{\mbtn{(2)}}\}=0\rightarrow\{\phi^{\mbtn{(4)}},\psi^{\mbtn{(2)}}\}=0\rightarrow\{\phi^{\mbtn{(3)}},\psi^{\mbtn{(1)}}\}=0.
\eqno{\mbox{(P2)}}
$$
\vs{12pt}\\
{\bf \boldmath I) Constraint Quantum System $\mcS^*_{\mbtn{III}}(\mbox{P}1)$}\footnote[8]{We shall denote the constraint quantum system by the process (P1) as $\mcS^*_{\mbtn{III}}(\mbox{P}1)$}\vs{12pt}\\
\ts{12pt}In the process (P1), $\phi^{\mbtn{(3)}},\phi^{\mbtn{(4)}}$ are first class and therefore would be considered as the generators of {\it gauge} transformations. Therefore, these are eliminated through imposing the following {\it gauge} fixing conditions\cite{A10,A11},
$$
\nu=0,\hs{36pt}\lambda=0.
\eqno{(3.45)}
$$
Then, $\mcS^*_{\mbtn{III}}(\mbox{P}1)$ is given as follows:
$$
\mcS^*_{\mbtn{III}}(\mbox{P}1)=(\mcC^*,\mcA(\mcC^*),H^*),
\eqno{(3.46a)}
$$
where
$$
\begin{array}{l}
\mcC^*=\{x^i,v_i|i=1,\cdots,N\},\vs{12pt}\\
\mbox{with}\e p^x_i=v_i,\e p_v^i=0,\vs{12pt}\\
\mcA(\mcC^*):\commut{x^i}{x^j}=0,\e \commut{x^i}{v_j}=i\hbar P_{ij}(x),\vs{12pt}\\
\ts{42pt}\commut{v_i}{v_j}=i\hbar\symmp{\mcG^{-1}(x)(G_{ik}(x)G_j(x)-G_i(x)G_{jk}(x))}{v_k},\vs{12pt}\\
H^*=\dps{\f12}\symmp{v_i}{v_i}+V(x)+\dps{\f{3\hbar^2}8}\mcG^{-2}(x)G_{ik}(x)G_k(x)G_{il}(x)G_l(x).
\end{array}
\eqno{(3.46b)}
$$
From (3.46b), thus, it is shown that $\mcS^*_{\mbtn{III}}(\mbox{P}1)$ is equivalent to $\mcS^*_{\mbtn{I}}$.
\vs{12pt}\\
{\bf \boldmath II) Constraint Quantum System $\mcS^*_{\mbtn{III}}(\mbox{P}2)$}\vs{12pt}\\
\ts{12pt}In the process (P2), $\psi^{\mbtn{(1)}},\phi^{\mbtn{(3)}}$ are first class. As well as in (P1), $\phi^{\mbtn{(3)}}$ is eliminated by imposing the {\it gauge} fixing condition $\nu=0$. On the other hand, 
$\psi^{\mbtn{(1)}}$ remains as the first-class operator. 
\\

Then, $\mcS^*_{\mbtn{III}}(\mbox{P}2)$ is constructed in the following way:
$$
\mcS^*_{\mbtn{III}}(\mbox{P}2)=(\mcC^*,\mcA(\mcC^*),H^*),
\eqno{(3.47a)}
$$
where
$$
\begin{array}{l}
\mcC^*=\{x^i,p^x_i|i=1,\cdots,N\},\vs{12pt}\\
\mbox{with}\vs{6pt}\\
v_i=\symmp{P_{ij}(x)}{p^x_j},\e p_v^i=0,\e \nu=p_{\nu}=0,\e \lambda=-\symmp{\mcG^{-1}(x)G_i(x)}{p^x_i},\e p_{\lambda}=0,\vs{12pt}\\
\mcA(\mcC^*):\commut{x^i}{p^x_j}=i\hbar\delta^i_j,\e \commut{x^i}{x^j}=\commut{p^x_i}{p^x_j}=0,\vs{12pt}\\
H^*=\dps{\f12}\symmp{\symmp{P_{ik}(x)}{p^x_i}}{\symmp{P_{jk}(x)}{p^x_j}}+V(x)+\dps{\f{3\hbar^2}8}\mcG^{-2}(x)G_{ik}(x)G_k(x)G_{il}(x)G_l(x).
\end{array}
\eqno{(3.47b)}
$$
Then, $\psi^{\mbtn{(1)}}=G(x)$ is the constant of motion, that is,
$$
\dot{G}(x)=(1/i\hbar)\commut{G(x)}{H^*}=0.
\eqno{(3.48)}
$$
From (3.47b) and (3.48), thus, it is shown that $\mcS^*_{\mbtn{III}}(\mbox{P}2)$ is equivalent to $\mcS^*_{\mbtn{II}}$.

\section{Discussion and Concluding remarks}

\ts{12pt}We have investigated the quantization of the dynamical system constrained to the submanifold $M^{N-1}$ specified by $G(x)=0$ in an $N$-dimensional Euclidean space $R^N$ through the POM with the constraint star-products. Then, we have obtained the following results:\vs{6pt}\\
\ts{12pt}(1) The projected quantum system $\mcS^{\mbtn{(2)}}$ includes two constraint quantum systems, one of which is $\mcS^*_{\mbtn{I}}$ with the {\it static} constraint, and the other of which, $\mcS^*_{\mbtn{II}}$ with the {\it dynamical} one, where
$$
\mcS^*_{\mbtn{I}}\subset\mcS^{\mbtn{(2)}}=\mcS^*_{\mbtn{II}}.
$$
Thus, it has been shown that $\mcS^*_{\mbtn{II}}$ is {\it complete}, and $\mcS^*_{\mbtn{I}}$, {\it incomplete}. \\
\ts{12pt}(2) The commutator algebras (3.21) and (3.29) in the constraint quantum systems are identical with the corresponding commutator ones obtained through the Dirac-bracket quantization formulation with the ordinaly second-order Lagrangians.\\
\ts{12pt}(3) The Hamiltonians in the constraint quantum systems contain the quantum correction terms due to the sequential projections of $\f12\symmp{v_i}{v_i}$ and $\symmp{\mu_{\mbtn{(3)}}}{\phi^{\mbtn{(3)}}}$\footnote[9]{In the operatorial approach, Lagrange multipliers are also operators at the first step.}, which are completely missed in the usual approach with Dirac-bracket quantization. In $\mcS^*_{\mbtn{II}}$, further, the Hamiltonian contains the additional correction terms associated to the re-ordering of operators in $\f12\symmp{v_i(x,p^x)}{v_i(x,p^x)}$, which will be obtained in the usual Dirac-bracket quantization procedure.\\
\ts{2pt}(4) Under the constraints $G(x)=0$ and $\dot{G}(x)=0$, there have been constructed two kinds of constraint quantum systems, $\mcS^*_{\mbtn{III}}(\mbox{P}1)$ and $\mcS^*_{\mbtn{III}}(\mbox{P}2)$. Then, it has been shown that $\mcS^*_{\mbtn{III}}(\mbox{P}1)$ is equivalent to $\mcS^*_{\mbtn{I}}$, and $\mcS^*_{\mbtn{III}}(\mbox{P}2)$, $\mcS^*_{\mbtn{II}}$\vs{6pt}\\
\ts{12pt}The approach based on first-order singular Lagrangians is considered to be one of the most available procedures for the investigation of noncommutative quantum system. Then, our next task is to extend our approach to noncommutative quantum systems. 
\vs{12pt}\\

\appendix

\ts{-24pt}{\bf \LARGE Appendix}

\section{Operations of ACCS}

\subsection{Operations of $\ximm{(1)}$,  $\pimm{(1)}$ on $\mcC^{\mbtn{(0)}}$}

$$
\begin{array}{lcl}
\ximm{(1)}_kx_i=\delta_{ki},&\hs{36pt}&\pimm{(1)}_kx^i=0,\vs{6pt}\\

\ximm{(1)}_kp^x_i=-\lambda G_{ki}(x),& &\pimm{(1)}_kp^x_i=0,\vs{6pt}\\

\ximm{(1)}_kv_i=0,& &\pimm{(1)}_kv_i=-\delta_{ki},\vs{6pt}\\

\ximm{(1)}_kp_v^i=\delta_{ki},& &\pimm{(1)}_kv_i=0,\vs{6pt}\\

\ximm{(1)}_k\lambda=0,& &\pimm{(1)}_k\lambda=0,\vs{6pt}\\

\ximm{(1)}_kp_{\lambda}=-G_k(x),& &\pimm{(1)}_kp_{\lambda}=0.

\end{array}
\eqno{(\mbox{A}1)}
$$

\subsection{Operations of $\ximm{(2)}$,  $\pimm{(2)}$ on $\mcC^{\mbtn{(1)}}$}

$$
\begin{array}{lcl}
\ximm{(2)}x_i&=&-\mcG^{-1}(x)G_i(x),\vs{12pt}\\

\ximm{(2)}p^x_i&=&-2\xipp{(2)}\mcG^{-1}(x)G_{ij}(x)G_j(x)+\symmp{\mcG^{-1}(x)G_{ij}(x)}{v_j}\vs{6pt}\\
& &+\lambda\mcG^{-1}(x)G_{ij}(x)G_j(x),\vs{12pt}\\

\ximm{(2)}v_i&=&-2\xipp{(2)}\mcG^{-1}(x)G_{ij}(x)G_j(x)+\symmp{\mcG^{-1}(x)G_{ij}(x)}{v_j},\vs{12pt}\\

\ximm{(2)}\lambda&=&0,\vs{12pt}\\

\ximm{(2)}p_{\lambda}&=&1,\vs{12pt}\\

\pimm{(2)}v_i&=&-G_i(x),\hs{36pt}(\pimm{(2)})^nv_i=0\hs{24pt}(n\geq 2),\vs{12pt}\\

\pimm{(2)}\lambda&=&-1,\vs{12pt}\\

& &\mbox{(the others)}=0.

\end{array}
\eqno{(\mbox{A}2)}
$$

\end{document}